\documentclass{article}
\usepackage{arxiv}
\pdfoutput=1

\usepackage[utf8]{inputenc} 
\usepackage[T1]{fontenc}    
\usepackage{hyperref}       
\usepackage{url}            
\usepackage{booktabs}       
\usepackage{amsfonts}       
\usepackage{nicefrac}       
\usepackage{microtype}      
\usepackage{array}
\usepackage[ruled,vlined]{algorithm2e}
\usepackage[inline]{enumitem}
\usepackage{lipsum}
\usepackage{graphicx}
\usepackage{placeins}
\usepackage{xcolor}
\usepackage{subcaption}
\usepackage{verbatim}
\graphicspath{ {./images/} }
\usepackage{listings}

\lstdefinelanguage{mlir}{
  keywords =
  {addf,affine,alloc,apply,call,memref,memref_vector_cast,dealloc,dma_start,dma_wait,if,for,scf,affine,memref,mulf,return,splat,tensor,to,load,store,gpu,subgroup_mma_load_matrix,subgroup_mma_store_matrix,subgroup_mma_compute,mma_matrix,launch,barrier,fpext,global,get_global},
	keywordstyle=\color{blue}\bfseries,
	ndkeywordstyle=\color{darkgray}\bfseries,
  identifierstyle=\color{black},
	sensitive=false,
	comment=[l]{//},
  commentstyle=\color{purple}\ttfamily,
}

\lstdefinelanguage{Shell}{
	keywords = {},
	keywordstyle=\color{blue}\bfseries,
	ndkeywordstyle=\color{darkgray}\bfseries,
  identifierstyle=\color{black},
	sensitive=false,
	comment=[l]{//},
	morecomment=[s]{/*}{*/},
  commentstyle=\color{purple}\ttfamily,
	stringstyle=\color{red}\ttfamily,
	morestring=[b]',
	morestring=[b]"
}

\lstdefinestyle{cmd} {
	language=Shell,
  columns=flexible,
  basicstyle=\small\ttfamily,
	literate={\$}{{\textcolor{blue}{\$}}}1
}

\lstdefinestyle{mlir} {
	language=mlir,
	  breaklines=true,                    
  basicstyle=\small\ttfamily,
  columns=flexible,
  mathescape
}

\usepackage{xifthen}

\title{High Performance GPU Code Generation for Matrix-Matrix Multiplication
using MLIR: Some Early Results}

\author{
  Navdeep Katel$^{1,2}$ \\
  \and
  \textbf{Vivek Khandelwal}$^{1}$\\
  \and
  \textbf{Uday Bondhugula}$^{1,2}$ \\
  \and
  $^1$ Dept of CSA, Indian Institute of Science \\
  Bengaluru, Karnataka 560012, India. \\
  \texttt{navdeepkatel@iisc.ac.in, vivekkhandel@iisc.ac.in, udayb@iisc.ac.in} \\
  \and
  $^2$ PolyMage Labs \\
  Entrepreneurship Centre, Indian Institute of Science \\
  Bengaluru, Karnataka 560012, India. \\
  \texttt{navdeep@polymagelabs.com, uday@polymagelabs.com} \\
}

\begin{document}

\newcolumntype{P}[1]{>{\centering\arraybackslash}p{#1}}
\newcolumntype{M}[1]{>{\centering\arraybackslash}m{#1}}
\newcolumntype{W}[1]{>{\hspace{0pt}}p{#1}}
\newcommand*\fmc[1]{\multicolumn{2}{|c}{\textbf{#1}}}
\newcommand*\bmc[1]{\multicolumn{2}{c|}{\textbf{#1}}}
\newcommand*\mmc[1]{\multicolumn{2}{|c|}{\textbf{#1}}}
\newcommand*\nmc[1]{\multicolumn{2}{c}{\textbf{#1}}}
\newcommand\Tstrut{\rule{0pt}{2.6ex}}       
\newcommand\Bstrut{\rule[-0.9ex]{0pt}{0pt}} 
\newcommand{\TBstrut}{\Tstrut\Bstrut} 

\date{}

\maketitle

\begin{abstract}

  This report presents some early results on code generation targeting tensor
  cores on NVIDIA GPUs using the MLIR compiler infrastructure. The 
  state-of-the-art in high-performance deep learning today is primarily driven 
  by highly tuned libraries. These libraries are often hand-optimized and tuned 
  by expert
  programmers using low-level abstractions with significant effort. A lot of 
  this
  effort may have to be repeated for similar hardware and future ones.  This
  process is thus not modular or reusable to the same extent that compiler
  infrastructure like LLVM are. Manual optimization typically does not use a
  standard intermediate representation (IR), although the optimizations 
  performed can be
  encoded as a sequence of transformation steps and customized passes on an IR.
  Hand tuning may also miss exploration of design points only reachable
  easily by automatic code generation. We believe that until the recent
  introduction of MLIR~\cite{mlir2020arxiv,mlir-web} (Multi-level intermediate
  representation), IR infrastructure was not geared to tackle the problem of
  automatic generation of domain-specific libraries in an effective manner. In 
  particular, it was hard to represent and transform compute abstractions at 
  high, middle, and low levels using a single IR.

  With suitable abstractions in MLIR, we build an experimental lowering pipeline 
  that is able to automatically generate code for matrix-matrix multiplication 
  on NVIDIA GPUs targeting its tensor cores. On a set of problem sizes we 
  evaluated,
  initial performance results show that we are able to attain performance that
  is 95-119\% and 80-160\% of CuBLAS~\cite{cublas} for \texttt{F32} and
  \texttt{F16} accumulate respectively on
  NVIDIA's Ampere~\cite{ampere-arch} microarchitecture-based Geforce 3090 RTX.
  We believe that these results could be used as motivation for
  further research and development on automatic code and library generation
  using IR infrastructure for similar specialized accelerators.
\end{abstract}

\section{Introduction}
  Deep learning and high-performance artificial intelligence in general heavily
  relies on high-performance computing today. The associated computing demands
  are being constantly met with innovations in computer hardware and
  micro-architecture, libraries, compilers, runtimes, and programming models.
  A significant amount of high-performance deep learning is currently powered by
  highly tuned libraries provided by hardware vendors: for example, CuDNN,
  CUBLAS, and MKL. Creating these libraries involves significant effort and
  expertise. This development process may have to be repeated with every major
  hardware or software release, and there are limits to what can be explored and
  optimized in an efficient manner.

  Matrix-matrix multiplication (matmul) as a computational kernel is at the
  heart of many deep learning frameworks based on
  Transformers~\cite{transformers} like BERT~\cite{bert}, and high-performance
  computing in general. It also serves as an excellent test case to guage what
  can be achieved. While automatic code generators' strength is often optimizing
  a composition of kernels as opposed to a well-known individual one, not being
  able to generate good code for a well-studied kernel can be a showstopper for
  the overall story of having automatic code generators tackle all peak
  performance code generation. In this report, we specifically target
  tensor cores on NVIDIA GPUs, which are specialized units for
  Matrix-Multiply-Accumulate (MMA) operations with typically 3-4x more
  throughput than normal CUDA cores.

  A few recent works have focused on GPU GEMM targeting tensor cores.
  Faingnaert et al.~\cite{juliaGemm} try to address the \textit{ two language
  problem } by creating a three-level API in Julia~\cite{Julia} that enables the
  user to write efficient GEMM kernels. Their primary focus is on
  developing an API flexible enough to cater to a wide variety of applications,
  it is not with a unified IR infrastructure with multiple levels of
  abstraction.  Bhaskaracharya et al.~\cite{voltaCodegen} take a polyhedral code generation approach to 
  generate code for Volta tensor cores.  They use schedule trees to represent the computation and use
  ISL~\cite{ISL} to generate CUDA code for it. They can generate code for matmul
  and fused operations like bias addition and ReLU, while achieving speed-ups up
  to 2.55$\times$. The work is specific to Volta, and includes some device
  specific specializations to achieve competitive performance. Tillet et
  al.~\cite{triton} recently presented Triton, an IR and an optimizing compiler
  for neural network computations.  The framework is based on the concept of a
  \textit{tile}, which is a statically shaped multi-dimensional array. The
  compiler is exposed by a python package which allows the user to write python
  code for which efficient machine code will be automatically generated. This
  work supports both CUDA and tensor core as well and achieves great
  performance.

  Our approach here is to use compiler intermediate representation (IR)
  infrastructure to facilitate high-performance code and library generation.
  We use matrix-matrix multiplication as the kernel to experiment and NVIDIA's
  tensor cores~\cite{tensor-cores} as the target.
  MLIR~\cite{mlir2020arxiv,mlir21cgo,mlir-web} is the compiler infrastructure we
  use here
  with the goal to make the whole process more modular, systematic, and
  automatic to a large extent. We show that by lowering the IR in small steps
  and applying the right set of IR transformations and optimizations, we can
  achieve performance that is comparable to those of hand-written libraries
  without actually writing any code by hand. While a previous
  work~\cite{bondhugula2020high} had done a similar study on a single core of a
  high-performance CPU, we are targeting a specialized accelerator here.

  Some of the contributions of this work include:
  \begin{itemize}
    \item the introduction of Warp Matrix Multiply Accumulate
      (WMMA)~\cite{wmma-ops-in-mlir} operations in an MLIR dialect along with
      their lowering to the LLVM/NVPTX backend,
    \item demonstrating how matmul on GPUs can be systematically and
      progressively code generated as a sequence of MLIR transformation and
    dialect lowering passes,
    \item building an end-to-end matmul code generation pipeline
      targeting tensor cores and preliminary results show that the performance
      obtained is on par with that of hand-tuned libraries and in some cases up
      to 1.60x better.
  \end{itemize}

  Our approach being an IR-based one can work with different programming models
  and languages if a lowering path from such models to MLIR exists.

\section{Background}
  \subsection{MLIR}
  Multi-Level Intermediate Representation (MLIR) \cite{mlir2020arxiv} aims to
  provide reusable, extensible compiler infrastructure
  and reduce the cost of building domain-specific compilers. MLIR can be used to
  serve multiple objectives such as:
  \begin{itemize}
    \item It can represent dataflow graphs (such as in TensorFlow), including
      dynamic shapes, variables, etc.
    \item It can be used to represent kernels for ML operations in a form
      suitable for optimization.
    \item It has the ability to host high-performance-computing-style loop
      optimizations across kernels (fusion, loop interchange, tiling,
      unroll-and-jam).
    \item It can represent target-specific operations, e.g.,
      accelerator-specific high-level operations.
    \item It can represent kernels at different levels of abstractions(dialects
      in MLIR), which can help perform transformations and optimizations not
      possible at a single level.
  \end{itemize}
  Some recent works that employ MLIR include \cite{gysi20arxiv,chelini21cgo}.

  The MLIR structure is made up of the following components:
  \begin{itemize}
    \item \textbf{Operations:} This is the basic unit of semantics in MLIR and
      is referred to as \textit{Op}. Everything from instructions to functions
      to modules is modeled as Ops in MLIR. Ops take and produce zero or more
      \textit{values}, called \textit{operands} and \textit{results}
      respectively.
    \item \textbf{Attributes:} It is structured compile-time static information
      e.g., integer constant values, string data, etc. Attributes are typed, and
      each Op instance has an open key-value dictionary from string names to
      attribute values.
    \item \textbf{Regions and blocks:} A \textit{region} contains a list of
      blocks, and a block contains a list of operations that may further contain
      regions. The blocks inside the region make a Control Flow Graph(CFG). Each
      block ends with a \textit{terminator} operation that may have
      \textit{successor} blocks to which the control flow may be transferred.
    \item \textbf{Dialects:} It is a logical grouping of Ops, attributes, and
      types under a unique namespace. Ops from different dialects can coexist at
      any level of the IR at any time. Dialects allow extensibility and provide
      flexibility that helps in performing specific optimizations and
      transformations. Affine, GPU, LLVM, and Linalg are some of the important
      dialects.
    \item \textbf{Functions and modules:} A module is an Op with a single region
      containing a single block and terminated by a dummy Op that does not
      transfer the control flow. A function is an Op with a single region, with
      arguments corresponding to function arguments.
  \end{itemize}

  Some MLIR dialects that we have used in our work are explained below:

  \begin{itemize}
    \item \textbf{Affine Dialect:} This dialect uses techniques from polyhedral
      compilation to make dependence analysis and loop transformations efficient
      and reliable. We have performed most of the optimizations and
      transformations at the level of affine dialect.
    \item \textbf{GPU Dialect:} The GPU dialect models the general GPU
      programming paradigm similar to CUDA or OpenCL in MLIR. Its goal is to
      provide abstractions to model GPU-specific operations and properties. It
      is largely meant to be vendor-agnostic. Some additional information can be
      found in
      \cite{herhut20gpu,zinenko19gpu} and the GPU dialect
      documentation~\cite{mlir-web}.
    \item \textbf{NVVM Dialect:} Since we are focusing on tensor core code
      generation, we use and extend the NVVM dialect.  This dialect provides
      operations that directly map to the NVPTX backend in LLVM.
    \item \textbf{LLVM Dialect:} The final stage of code generation involves
      lowering to LLVM IR, from where the LLVM backend takes control and
      generates the target code. To model LLVM IR, this dialect is used. This is
      the lowest level of abstraction present in MLIR.
  \end{itemize}

  \subsection{GPU Background}
  GPUs are general-purpose massively parallel computing devices. The memory and
  the compute hierarchy play an essential role in optimizing any application and
  thus achieving high performance. We can abstract the GPU memories into a
  4-level hierarchy, global memory, L2-cache, a configurable L1-cache/shared
  memory, and registers. The processors on GPUs can also be abstracted into a
  two-level hierarchy, the Streaming Multiprocessors (SMs) and computing cores
  inside the SM. The computing cores are often known as CUDA cores. Besides CUDA
  cores, special units like tensor cores are also present in newer GPUs at same
  level in the compute hierarchy. Each SM is further divided into processing
  blocks which have their respective warp schedulers. The programming model of
  GPUs is also structured to match the processor hierarchy present. A
  \textit{thread} is a single execution entity on a GPU that can execute in
  parallel with other threads. These threads are clubbed into groups of 32,
  called a \textit{warp}.  Warps execute in a lockstep manner on the computing
  cores present on the SM. A \textit{warp scheduler} chooses a warp that is
  ready to execute and dispatches it to the computing cores. When a warp
  encounters a data dependency, it stalls, and the warp scheduler chooses
  another warp that is ready for execution. Depending on the number of
  processing blocks present on the SM, multiple warps may be executing in
  parallel. So, in general, more warps help one achieve:
  \begin{enumerate*}[label=(\roman*), itemjoin={{, }}, itemjoin*={{, and }}]
    \item parallelism at the warp level,
    \item better latency hiding,
    \item and better utilization of the underlying resources.
  \end{enumerate*}
  Now, these warps are further grouped into a \textit{thread block}. There can
  be several thread blocks that may be executing in parallel on the GPU. A
  thread block is bound to an SM. It cannot change the SM during its execution
  lifetime and must complete its execution on the same SM and release all the
  resources allocated to it upon completion. Threads in the same warp can
  exchange data using warp level shuffle
  instructions~\cite{warp-level-primitives}. All the threads in a thread block
  can communicate using low latency \textit{shared memory}, and threads in
  different thread blocks need to communicate using the high latency
  \textit{global memory}. Synchronization primitives are present at thread block
  and warp level. A Synchronization will ensure that none of the threads in a
  thread block or a warp, depending on the type of synchronization used, proceed
  to the next instruction until all of them have arrived at the synchronization
  point.  Synchronization becomes necessary while using shared memory in a case
  where data is first written to shared memory and then read by all threads. All
  threads must be synchronized before reading from and writing into the shared
  memory buffers to ensure correctness.

  \subsection{Tensor Cores}
  \label{sec:tensor_core_background}
  Tensor Cores \cite{tensor-cores} are programmable
  matrix-multiply-and-accumulate (MMA) units present on NVIDIA GPUs. First
  introduced in Volta architecture, they are present on Turing and Ampere as
  well. Significantly more throughput than CUDA cores makes them excellent for
  accelerating deep learning workloads. They perform MMA operation represented
  as, \textbf{D} = \textbf{A} * \textbf{B} + \textbf{C}, where the operations
  size is \textbf{4$\times$4$\times$4} on Turing and Volta while it is
  \textbf{8$\times$4$\times$8} on Ampere. Tensor cores execute \textit{warp
  synchronous} instructions like \texttt{HMMA} to perform the MMA operation.
  \texttt{Warp synchronous} means that all the threads in a warp cooperatively
  execute these special instructions in order to produce the output of the MMA
  operation. Due to this \texttt{warp synchronous} nature of the tensor cores
  instructions, it becomes necessary to write or generate code at the warp level
  rather than the thread level when programming tensor cores. Tensor cores
  originally supported only \texttt{FP16} for input and \texttt{FP16} or
  \texttt{FP32} for accumulation and output. But now they support a variety of
  formats for both input and output like \texttt{TF32}, \texttt{BF16},
  \texttt{INT8}, and \texttt{INT4}. \texttt{TF32} has the same range as
  \texttt{FP32}, and the same precision as \texttt{FP16}, but is represented in
  19-bits. It can be used at places where precision is of less importance. To
  use this mode the inputs have to be in \texttt{FP32}, they will be internally
  converted to \texttt{TF32}, accumulation will done in \texttt{FP32}, and the
  output will be produced in \texttt{FP32} as well. This provides a speed-up
  over normal \texttt{FP32} mode on CUDA cores. \texttt{BF16} offers same range
  as \texttt{FP32}, and precision which is less than \texttt{FP16}. Tensor cores
  offer the same speed in both \texttt{BF16} and \texttt{FP16} modes, while both
  are faster than \texttt{TF32}. The integer types are meant to be used in post
  training quantization~\cite{post-training-quantization}.
  
  When it comes to programmability, there are three ways to leverage tensor
  cores:
   \begin{enumerate*}[label=(\roman*), itemjoin={{, }}, itemjoin*={{, and }}]
    \item use a high level library like cuBLAS,
    \item program using a high level C++ WMMA API~\cite{wmma-api-in-cuda} in
      CUDA, or
    \item program them explicitly using assembly level instructions.
  \end{enumerate*}
  
  \begin{center}
    \footnotesize
    \begin{tabular}[!h]{ M{2cm} M{3cm} M{3cm} M{3cm} M{3cm} }
      \hline
      & {\parbox{2cm}{\centering \TBstrut \textbf{Performance}\TBstrut}} &
      {\parbox{3cm}{\centering \TBstrut \textbf{Shared memory bank conflicts}\TBstrut}} &
      {\parbox{3cm}{\centering \TBstrut \textbf{Ease of use}\TBstrut}} &
      {\parbox{3cm}{\centering \TBstrut \textbf{Support for operator fusion}\TBstrut}} \\
      \hline
      \textbf{High-level libraries} & Best & minimal & Accessible via function
	calls & limited \\
      \hline
      \textbf{WMMA API} & Competitive in most cases & higher & Programming effort
      required & good \\
      \hline
      \textbf{Assembly} & May be competitive in all cases & minimal & 
      Significant
      programmer effort required & good \\
      \hline
    \end{tabular}
    \captionof{table}{Comparison of different approaches to program tensor cores.}
    \label{table:approaches-to-program-tcs}
  \end{center}

  While cuBLAS can be used using mere functions calls, using the other two
  methods, require significant programming efforts. WMMA API provides larger
  matrix operations (16$\times$16$\times$16, 32$\times$8$\times$16), and utility
  functions to load and store operand matrices. The task of converting these API
  functions to GPU microarchitecture specific assembly instructions is also
  offloaded to NVIDIA's proprietary compile. The matrices loaded using WMMA API
  have and opaque layout once they are loaded into the registers, i.e., which
  thread is holding which element of the loaded matrix (thread-data mapping) is
  not known.  Due to this opaque nature, some additional steps are needed when
  doing fusion with operations like bias-addition, which need to be aware of the
  thread-data mapping. Programming tensor cores explicitly using assembly
  instructions is even more challenging because the programmer has to take care
  of complexities such as thread data mappings in registers and data movement
  between shared memory and registers. The approaches are summarized in
  Table~\ref{table:approaches-to-program-tcs}.

  The \texttt{NVPTX} backend in LLVM exposes WMMA API functions as \texttt{intrinsics}.
  This makes it possible to program the tensor cores via MLIR. These intrinsics
  map one-to-one with the WMMA API functions, and exhibit identical behaviour in
  terms of programming and usage.

\section{Design}
  \label{sec:design}
  \begin{figure}[!h]
    \centering
    \includegraphics[width=\columnwidth]{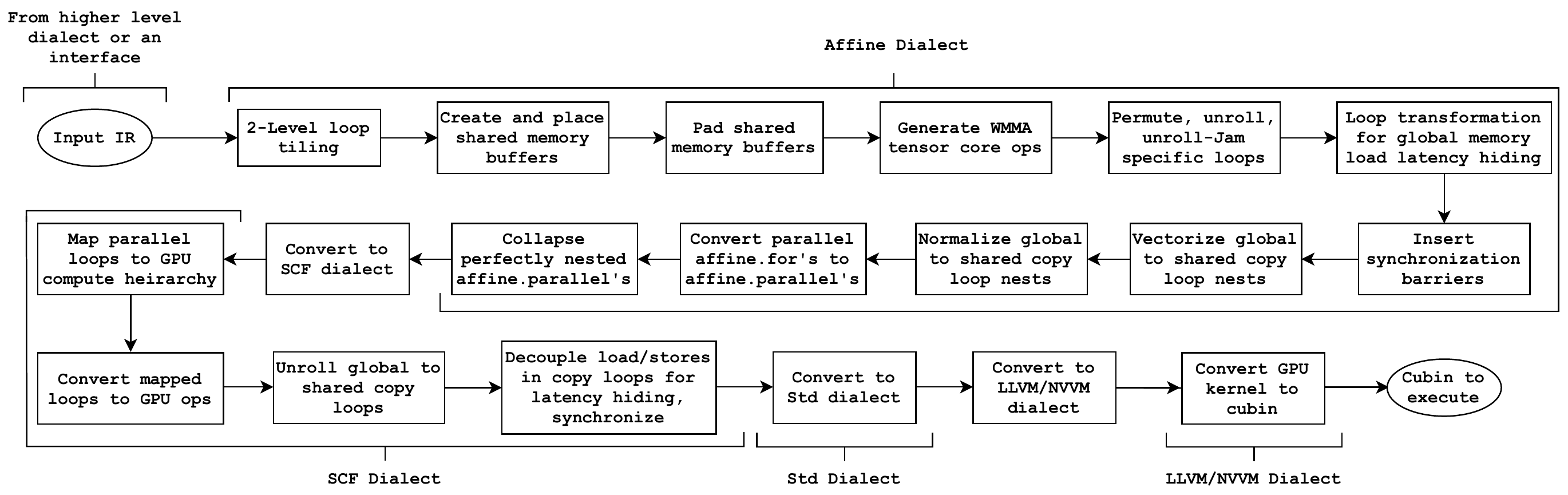}
    \caption{Lowering path for matmul to target GPU tensor 
  cores.~\label{fig:lowering-path}}
  \end{figure}
  In this section we present the design of our pipeline. Our pipeline is based
  on a series of optimizations and transformations which make up the recipe for
  fast matmul on GPUs. The recipe we use is close to what is highlighted in some
  prior works[\cite{cutlass},\cite{juliaGemm},\cite{voltaCodegen}]. The part
  that is common in all of them is two-level blocking to maximize the reuse at
  different levels of the memory hierarchy. The general recipe is described in
  Algorithm~\ref{algo:matmul-recipe}.

  Prior to our work, some support was present in MLIR, which we have reused in 
  our pipeline, but some core components were missing. Mainly, the ops required
  to program the tensor cores using the WMMA API were absent from MLIR, and we
  were the ones to introduce these ops. We make changes and additions to the
  existing MLIR infrastructure wherever necessary.

  Figure~\ref{fig:lowering-path} shows the lowering path that we take, which is
  based on Algorithm~\ref{algo:matmul-recipe}. Although there can be different
  lowering paths to achieve the same goal, we argue that the one going through
  affine dialect should be chosen as long as the target kernel to be generated
  is affine. This could help in many things such as creation and placement of
  fast memory buffers, loop-tiling, unroll-jam, vectorization, detection of
  parallel loops, and placement of synchronization barriers too.
  
  \begin{algorithm}[!h]
    \SetAlgoLined
    \SetKw{KwBy}{step}
      Global memory: A[\textit{M}][\textit{K}] B[\textit{K}][\textit{N}], C[\textit{M}][\textit{N}]\;
      Shared memory: a\_smem[\textit{tbm}][\textit{tbk}],
      b\_smem[\textit{tbk}][\textit{tbn}]\;
      Registers: c\_reg[\textit{wm}][\textit{wn}], a\_reg[\textit{wm}],
      b\_reg[\textit{wn}]\;
      \For{$threadBlockK\gets0$ \KwTo $M$ \KwBy $tbk$}{
	\textit{\_\_syncthreads()}\;
	All threads load \textit{tbm} $\times$ \textit{tbk} block form A to
	a\_smem\;
	All threads load \textit{tbk} $\times$ \textit{tbn} block form B to
	b\_smem\;
	Warp loads \textit{wm} $\times$ \textit{wn} block form C to
	c\_reg[\textit{wm}][\textit{wn}]\;
	\textit{\_\_syncthreads()}\;
	\CommentSty{// \texttt{wmmaM}, \texttt{wmmaN}, and \texttt{wmmaK} represents the
	WMMA intrinsic size}\;
	\For{$warpK\gets0$ \KwTo $tbk$ \KwBy $wmmaK$}{
	  \For{$warpM\gets0$ \KwTo $wm$ \KwBy $wmmaM$}{
	    Warp loads a fragment of \textit{A} form a\_smem\ into
	    a\_reg[\textit{warpM}]\;
	    \For{$warpN\gets0$ \KwTo $wn$ \KwBy $wmmaN$}{
	      Warp loads a fragment of \textit{B} form b\_smem\ into
	      b\_reg[\textit{warpM}]\;
	      c\_reg[\textit{warpM}][\textit{warpN}] $\mathrel{+}=$
	      a\_reg[\textit{warpM}] $\times$
	      b\_reg[\textit{warpN}]\;
	    }
	  }
	}
	Warp stores c\_reg[\textit{wm}][\textit{wn}] to the respective block in C\;
      }
      \caption{Two-level tiled tensor core Matmul.~\label{algo:matmul-recipe}}
  \end{algorithm}

  Although not highlighted in Algorithm~\ref{algo:matmul-recipe} for brevity, It
  is worth noting that high performance can only be achieved using a set of some
  more optimizations, which include
  \begin{enumerate*}[label=(\roman*), itemjoin={{, }}, itemjoin*={{, and }}]
    \item padding in shared memory buffers to reduce bank conflicts
    \item register tiling or warp tiling
    \item load-store vectorization
    \item global memory load latency hiding.
  \end{enumerate*}
  Now, we will describe our lowering pipeline in some detail, discussing
  how the major optimizations are enabled. We skip over the trivial parts.

  \subsection{A Starting Point}
  The starting point for our code generation approach is a high-level op like
  \texttt{lmhlo.dot} or \texttt{linalg.matmul} or simply an IR that is generated
  from a user-facing programming model targeting a linear algebra dialect. In
  the former case, we can lower the op to a three-loop affine matmul, while we
  can generate the three-loop affine matmul in the latter. Our starting point is
  shown in Listing~\ref{lst:naive-affine}.

  \begin{figure}
  \begin{lstlisting}[style=mlir, caption=Naive affine matmul., captionpos=b,
  label=lst:naive-affine]
  affine.for %i = 0 to %M {
    affine.for %j = 0 to %N {
      affine.for %k = 0 to %K {
        %a = affine.load %A[%i, %k] : memref<8192x8192xf16>
        %b = affine.load %B[%k, %j] : memref<8192x8192xf16>
        %c = affine.load %C[%i, %j] : memref<8192x8192xf32>
        %aq = fpext %a : f16 to f32
        %bq = fpext %b : f16 to f32
        %q = mulf %aq, %bq : f32
        %co = addf %c, %q : f32
        affine.store %co, %C[%i, %j] : memref<8192x8192xf32>
      }
    }
  }
  \end{lstlisting}
  \end{figure}

  \subsection{Tiling for Locality and Parallelism}
  \label{sec:tiling}
    It is well known that tiling helps to exploit data reuse and significantly
    improves performance if the correct parameters are chosen for tiling.
    Two-level tiling is essential to achieve optimal performance in GPUs. The
    first level of tiles are mapped to different thread blocks. Each thread
    block copies its tile for \texttt{A} and \texttt{B} from global memory to
    shared memory, thus preventing multiple high latency trips to global memory.
    As the partitioned tiles are mapped to different thread blocks, they can be
    computed in parallel on different SMs. The second level of tiling is done to
    exploit reuse in the registers and promote warp-level parallelism. The
    thread block level tile is partitioned between the warps, and each warp only
    works on the part part of the tile mapped to it. This step gives us the two 
    level tiled structure to proceed with.

  \subsection{Creating and Placing Shared Memory Buffers}
    After tiling is done, the next step is to create and place shared memory
    buffers at the right loop depth. We use \texttt{affineDataCopyGenerate}
    utility to generate the copies for matrices \texttt{A} and \texttt{B}. We
    take a slightly different approach here than some previous works.
    We create shared memory buffers only for matrices \texttt{A} and \texttt{B}.
    Since \texttt{C} is only loaded once per warp, we stream it directly into
    the registers from the global memory. Faingnaert et al~\cite{juliaGemm}
    stream \texttt{C} from global memory to shared memory and then from shared
    memory to registers. The rationale for streaming \texttt{C} via shared
    memory is the prevention of random accesses to global memory and perhaps
    promoting coalesced accesses in global memory, which could be more
    efficient. We speculate that this may not always be the case, especially for
    large problem sizes, as \texttt{C} tile is only loaded once per warp.

    Additionally, this approach would also require the use of dynamically
    allocated shared memory, as the optimal tile size for \texttt{C} could
    easily exhaust the 48 KB static shared memory limit on some devices. So the buffer to hold
    the tiles must be allocated dynamically and must be reused for storing tiles
    of all three operands. The support to allocate shared memory dynamically is
    currently absent from MLIR. So we restrict ourselves to statically allocated
    shared memory to avoid additional complexities in the code generator which
    may not be worth the effort. Even without doing this we already reach close
    to hand-tuned libraries in most cases.
    
    Creating the shared memory buffers is one part of the story while ensuring
    shared memory accesses with minimal bank conflicts is another.  Bank
    conflicts can reduce the throughput of shared memory by a significant
    amount. A general technique to avoid bank conflicts is to pad the shared
    memory buffers in the leading dimension. We achieve the same thing by
    changing the \(leadingDimension\) of the shared memory buffer that was
    generated by \texttt{affineDataCopyGenerate} to \(leadingDimension +
    paddingFactor\). Doing this will change the underlying layout map of the
    shared memory buffer to account for the change in the leading dimension, and
    the rest of the IR need not be changed. It is also worth noting that we can
    try out different padding factors here and see what performs the best, but
    the padding factor must be a multiple of 8, i.e., 128-bits for \texttt{F16}
    elements. This is because of an alignment requirement by the WMMA API.

  \subsection{Generating WMMA Ops and More}
    Now that we have all the basic things that we need, we can proceed towards
    the generation of \texttt{gpu.subgroup\_mma} ops. The WMMA ops are of
    different sizes and we use the 16$\times$16$\times$16 version of the ops in
    this work. The ops we generate here should replace the scalar ops that are
    already present, and the loop steps of the corresponding loops must be
    adjusted accordingly.
    \begin{lstlisting}[style=mlir, caption=Tiled and padded affine matmul with
    WMMA ops., captionpos=b, label=lst:affine-specialized]
    #map0 = affine_map<(d0) -> (d0)>
    #map1 = affine_map<(d0) -> (d0 + 64)>
    #map2 = affine_map<(d0) -> (d0 + 128)> 
    module {
      // Shared memory buffers for A and B.
      memref.global "private" @b_smem_global : memref<64x136xf16, 3>
      memref.global "private" @a_smem_global : memref<128x72xf16, 3>
      func @main() {
        ...
	affine.for %i = 0 to 8192 step 128 {
	  affine.for %j = 0 to 8192 step 128 {
	    // References to shared memory buffers.
    	    %b_smem = memref.get_global @b_smem_global : memref<64x136xf16, 3>
    	    %a_smem = memref.get_global @a_smem_global : memref<128x72xf16, 3>
	    // Main k-loop.
    	    affine.for %k = 0 to 8192 step 64 {
	      // Copy loop for B.
    	      affine.for %copykk = #map0(%k) to #map1(%k) {
    	        affine.for %copyjj = #map0(%j) to #map2(%j) {
    	          %11 = affine.load %B[%copykk, %copyjj] : memref<8192x8192xf16>
    	          affine.store %11, %b_smem[%copykk - %k, %copyjj - %j] : memref<64x136xf16, 3>
    	        }
    	      }
	      // Copy loop for A.
    	      affine.for %copyii = #map0(%i) to #map2(%i) {
    	        affine.for %copykk = #map0(%k) to #map1(%k) {
    	          %11 = affine.load %A[%copyii, %copykk] : memref<8192x8192xf16>
    	          affine.store %11, %a_smem[%copyii - %i, %copykk - %k] : memref<128x72xf16, 3>
    	        }
    	      }
    	      affine.for %ii = 0 to 128 step 64 {
    	        affine.for %jj = 0 to 128 step 32 {
    	          affine.for %kk = 0 to 64 step 32 {
                    affine.for %kkk = 0 to 32 step 16 {
                      affine.for %iii = 0 to 64 step 16 {
                        affine.for %jjj = 0 to 32 step 16 {
                          ...
                          %a = gpu.subgroup_mma_load_matrix %a_smem[%11, %12] {leadDimension = 72 : index} : memref<128x72xf16, 3> -> !gpu.mma_matrix<16x16xf16, "AOp">
                          %b = gpu.subgroup_mma_load_matrix %b_smem[%12, %14] {leadDimension = 136 : index} : memref<64x136xf16, 3> -> !gpu.mma_matrix<16x16xf16, "BOp">
                          %c = gpu.subgroup_mma_load_matrix %C[%16, %17] {leadDimension = 8192 : index} : memref<8192x8192xf32> -> !gpu.mma_matrix<16x16xf32, "COp">
                          %res = gpu.subgroup_mma_compute %a, %b, %c : !gpu.mma_matrix<16x16xf16, "AOp">, !gpu.mma_matrix<16x16xf16, "BOp"> -> !gpu.mma_matrix<16x16xf32, "COp">
                          gpu.subgroup_mma_store_matrix %res, %C[%16, %17] {leadDimension = 8192 : index} : !gpu.mma_matrix<16x16xf32, "COp">, memref<8192x8192xf32>
                        }
                      }
                    }
    	          }
    	        }
    	      }
    	    } 
    	  }
    	}
      }
    }
    \end{lstlisting}

    \TBstrut
    Now that we have generated the WMMA operations, we do the following IR
    transformations:
    \begin{itemize}
      \item permute the outermost six loops to go from (\(\texttt{i},
	\texttt{j}, \texttt{k}, \texttt{ii}, \texttt{jj}, \texttt{kk}\)) order
	to (\(\texttt{i}, \texttt{j}, \texttt{ii}, \texttt{jj}, \texttt{k},
	\texttt{kk}\)) order. This later helps in mapping compute loops to GPU
	compute hierarchy.  Additionally, it also helps in moving invariant
	load-store operations on \texttt{C} to the most outermost position
	possible.
      \item permute the innermost three loops to go form (\(\texttt{i},
	\texttt{j}, \texttt{k}\)) to (\(\texttt{k}, \texttt{i}, \texttt{j}\)).
	This represents warp level MMA operation as an outer product and
	enhances ILP, as pointed out by Bhaskaracharya et
	al.~\cite{voltaCodegen}.
      \item fully unroll the innermost three loops.
    \end{itemize}

    Listing~\ref{lst:affine-specialized} shows the IR that we get after the
    creation of WMMA ops. We should note the step of the innermost loops being
    adjusted here. The listing further shows the loop nest in a permutation we
    desire.  The two outermost loops will later be mapped to the thread blocks
    in a grid, and the following two loops will be mapped to the warps. The next
    two loops are the k-loops corresponding to the thread block and then the
    warp. After unrolling, we make two observations
    \begin{enumerate*}[label=(\roman*), itemjoin={{, }}, itemjoin*={{, and }}]
    \item the operations on \texttt{C} matrix now become independent of the two
      surrounding loops, and hence we hoist the operations on \texttt{C} now
      to the outermost k-loop. In this way, we prevent repetitive loads and
      stores to \texttt{C} in the global memory and only do them when the
      processing of a thread block tile begins and ends
    \item unrolling these loops completely reveal all loads on \texttt{A} and
      \texttt{B}. Some of these loads are the same along the k-dimension, and by
      applying CSE, we can completely remove the redundant loads and achieve
      unroll-jam kind of effect.
    \end{enumerate*}
    
    \begin{lstlisting}[style=mlir, caption=Affine matmul after loop unrolling and
    invariant load-store hoisting., captionpos=b,
    label=lst:unrolled-affine-matmul-wmma, escapechar=|]
    ...
    #map0 = affine_map<(d0, d1) -> (d0 + d1)>
    ...
    // Thread block `i` loop.
    affine.for %i = 0 to 8192 step 128 {
      // Thread block `j` loop.
      affine.for %j = 0 to 8192 step 128 {
        %b_smem = memref.get_global @b_smem_global : memref<64x136xf16, 3>
        %a_smem = memref.get_global @a_smem_global : memref<128x72xf16, 3>
        // Warp `i` loop.
        affine.for %ii = 0 to 128 step 64 {
          // Warp `j` loop.
          affine.for %jj = 0 to 128 step 32 {
            // Hoisted loads on C.
            %11 = affine.apply #map0(%i, %ii)
            %12 = affine.apply #map0(%j, %jj)
            %c_reg_0 = gpu.subgroup_mma_load_matrix %C[%11, %12] {leadDimension = 8192 : index} : memref<8192x8192xf32> -> !gpu.mma_matrix<16x16xf32, "COp">
            ...
            // Main `k`-loop with loaded C operand as iter_args.
            %res:8 = affine.for %k = 0 to 8192 step 64 iter_args(%c_in_0 = %c_reg_0, %c_in_1 = %c_reg_1 ...) -> (!gpu.mma_matrix<16x16xf32, "COp">, !gpu.mma_matrix<16x16xf32, "COp">) { |\label{line:global-k}|
              ...
              %a = gpu.subgroup_mma_load_matrix %a_smem[%ii, %c_in_0] {leadDimension = 72 : index} : memref<128x72xf16, 3> -> !gpu.mma_matrix<16x16xf16, "AOp">
              %b = gpu.subgroup_mma_load_matrix %b_smem[%c_in_0, %jj] {leadDimension = 136 : index} : memref<64x136xf16, 3> -> !gpu.mma_matrix<16x16xf16, "BOp">
              %c_res = gpu.subgroup_mma_compute %a, %b, %c_in_0 : !gpu.mma_matrix<16x16xf16, "AOp">, !gpu.mma_matrix<16x16xf16, "BOp"> -> !gpu.mma_matrix<16x16xf32, "COp">
              ...
	      // Main `k`-loop yielding the results of the current iteration.
              affine.yield %104, %107 ... : !gpu.mma_matrix<16x16xf32, "COp">, !gpu.mma_matrix<16x16xf32, "COp">...
            }
            // Hoisted stores on C.
            gpu.subgroup_mma_store_matrix %res#0, %C[%11, %12] {leadDimension = 8192 : index} : !gpu.mma_matrix<16x16xf32, "COp">, memref<8192x8192xf32>
            ...
          }
        }
      }
    }
    \end{lstlisting}
    
    The loop structure just after the said optimizations is shown in
    Listing~\ref{lst:unrolled-affine-matmul-wmma}. It must be noted how the loop
    structure has changed after the movement of invariant load-store pairs of
    \texttt{C} matrix. The \texttt{affine.for} operation in line~\ref{line:global-k}
    represents the main-k loop, and is now modified to take the loaded \texttt{C}
    operands as loop \texttt{iter\_args}.  These will serve as accumulators for
    the multiplications happening in this loop. After every iteration, this loop
    \texttt{yields} accumulated products, which are passed as \texttt{iter\_args}
    to the next iteration. These \texttt{iter\_args} reside in the registers and
    are reused across different iterations of the main k-loop.

  \subsection{Global Memory Load Latency Hiding}
  \label{sec:latency-hiding}
  With the introduction of \texttt{gpu.subgroup\_mma} ops and some other
  optimizations in the previous section, we are moving towards the structure
  that we will have in the final IR. We focus on doing as many optimizations as
  possible without having any GPU-specific information in the affine dialect
  itself. In the IR we have currently, we cannot start computation until we have
  loaded the shared memory tiles of \texttt{A} and \texttt{B}. Global memory
  loads are one of the most expensive operations in terms of latency, so
  eliminating the long wait times on the operands is very important. We do so by
  splitting the main k-loop or the thread block k-loop by taking out the copy of
  \texttt{A} and \texttt{B} for iteration \texttt{0} and compute for iteration
  \texttt{n - 1}. The copy is placed just before the k-loop, and compute is
  placed just after it. The indices for the computation being performed in that
  loop also need to be shifted to move one iteration ahead. As a result, compute
  happens on data already available in shared memory with loads for the next
  iteration having been issued. We show the structure of the IR at this stage in
  Listing~\ref{lst:affine-delayed}.

  \begin{lstlisting}[style=mlir, caption=WMMA affine matmul with shifted
  k-loop., captionpos=b, label=lst:affine-delayed]
    #map4 = affine_map<(d0) -> (d0)>
    #map5 = affine_map<(d0) -> (d0 + 128)>
    #map6 = affine_map<(d0) -> (d0 + 64)>
    // Peeled copy loops for iteration 0 of k-loop.
    affine.for %copyk = 0 to 64 {
      affine.for %copyj = #map4(%j) to #map5(%j) {
        %35 = affine.load %B[%copyk, %copyj] : memref<8192x8192xf16>
        affine.store %35, %b_smem[%copyk, %copyj - %j] : memref<64x136xf16, 3>
      }
    }
    affine.for %copyi = #map4(%i) to #map5(%i) {
      affine.for %copyk = 0 to 64 {
        %35 = affine.load %A[%copyi, %copyk] : memref<8192x8192xf16>
        affine.store %35, %a_smem[%copyi - %i, %copyk] : memref<128x72xf16, 3>
      }
    }
    // Main k-loop.
    affine.for %k = 0 to 8128 step 64 {
      // Copy loops for iteration `%k + 1` of k-loop.
      affine.for %copyk = #map6(%k) to #map5(%k) {
        affine.for %copyj = #map4(%j) to #map5(%j) {
          %36 = affine.load %B[%copyk, %copyj] : memref<8192x8192xf16>
          affine.store %36, %b_smem[%copyk - %k - 64, %copyj - %j] : memref<64x136xf16, 3>
        }
      }
      affine.for %copyi = #map4(%i) to #map5(%i) {
        affine.for %copyk = #map6(%k) to #map5(%k) {
          %36 = affine.load %A[%copyi, %copyk] : memref<8192x8192xf16>
          affine.store %36, %a_smem[%copyi - %i, %copyk - %k - 64] : memref<128x72xf16, 3>
        }
      }
      affine.for %kk= 0 to 64 step 32 {
        ...
      }
    }
    // Peeled compute loop for the last iteration of k-loop.
    affine.for %arg4 = 8128 to 8192 step 64 {
      ...
    }
  \end{lstlisting}
    While this lays the groundwork for latency hiding, to see this in action, we
    need to decouple the stores into shared memory from the loads on global
    memory for the copy loops inside the thread block k-loop. This is required
    both for the correctness and functioning of the optimization. For this, we
    unroll the copy loops and delay stores to be the trailing operations in the
    outer k-loop. We delay this optimization to a further point in the pipeline
    as some GPU-specific information is required to enable it.

  \subsection{Inserting Synchronization Barriers}
    We have finished generating most of the parts in the IR, which may require
    synchronization barriers. The shared memory buffers will be read and written
    by all the threads in the thread block, so it is essential to synchronize
    before and after writing into these buffers. In general, this process can
    also be automated using memory-based dependence analysis. However, we place
    these synchronization barriers using the aforementioned static information
    about the copy loops for our current purposes.

  \subsection{ Global to Shared Copy Vectorization}
    While latency hiding will do its bit, it cannot make the actual copy go any
    faster. It is well known that vector load
    store instructions~\cite{vectorized-memory-accesses}perform better than their scalar 
    counterparts because it
    reduces the number of memory transactions and often results in better
    utilization of the available bandwidth.

    We use the vectorization utilities already present in MLIRX~\cite{mlirx}. We
    call this utility on the global to shared memory copies. We can try out
    different widths for the vector widths using this utility. We tried out 32,
    64 and 128 bit wide vectors and found out the 128-bit wide vectors to work
    the best. Vectorized copy loops are shown in Listing~\ref{lst:vector-copy}.
    \begin{lstlisting}[style=mlir, caption=Vectorized copy loops., captionpos=b,
    label=lst:vector-copy]
    ...
    #map4 = affine_map<(d0) -> (d0)>
    #map5 = affine_map<(d0) -> (d0 + 128)>
    #map6 = affine_map<(d0) -> (d0 + 64)>
    ...
    // Cast operations for global memory memrefs.
    %a_cast = memref.vector_cast %A : memref<8192x8192xf16> to memref<8192x1024xvector<8xf16>>
    %b_cast = memref.vector_cast %B : memref<8192x8192xf16> to memref<8192x1024xvector<8xf16>>
    // Cast operations for shared memory memrefs.
    %b_smem_cast = memref.vector_cast %b_smem : memref<64x72xf16, 3> to memref<64x9xvector<8xf16>, 3>
    %a_smem_cast = memref.vector_cast %a_smem : memref<128x72xf16, 3> to memref<128x9xvector<8xf16>, 3>
    // Vectorized copy loops.
    affine.for %copyk = #map6(%k) to #map5(%k) {
      affine.for %copyj = #map4(%j) to #map5(%j) step 8 {
        %135 = affine.load %b_cast[%copyk, %copyj floordiv 8] : memref<8192x1024xvector<8xf16>>
        affine.store %135, %b_smem_cast[%copyk - %k - 64, (%copyj - %j) floordiv 8] : memref<64x17xvector<8xf16>, 3>
      }
    }
    affine.for %copyi = #map4(%i) to #map5(%i) {
      affine.for %copyk = #map6(%k) to #map5(%k) step 8 {
        %135 = affine.load %a_cast[%copyi, %copyk floordiv 8] : memref<8192x1024xvector<8xf16>>
        affine.store %135, %a_smem_cast[%copyi - %i, (%copyk - %k) floordiv 8 - 8] : memref<128x9xvector<8xf16>, 3>
      }
    }
    \end{lstlisting}

  \subsection{Extracting Parallel Loops.}
    This is the last step that we take while still in the affine dialect. We use
    \texttt{isLoopParallel} utility present in MLIR to find all the parallel
    loops and then convert them to parallel loops using
    \texttt{affineParallelize}. These parallel loops are later processed and
    mapped to the GPU processor hierarchy, while the sequential loops are the
    only ones which will remain in the kernel.

  \subsection{Mapping to GPU Compute Hierarchy}
    The previous step was the last one in the affine dialect, and we immediately
    convert to the SCF dialect after it. Starting in the SCF dialect, the first
    thing that we do is map the parallel loops to the GPU compute hierarchy. The
    existing utilities and passes in MLIR for mapping do not support mapping
    loops to individual warps, which is required in our case. We extend the
    utilities and passes to add this support for matmul. Ideally, we should
    generalize the passes and utilities used in this step to handle a wide
    variety of loop nests, we leave it as future work. It is worth pointing
    out at this point that we take all the measures necessary to ensure coalesced global memory accesses~\cite{global-memory-coalescing} which are crucial for effective
    bandwidth utilization and faster copies from global to shared memory. After
    mapping is completed, the two outermost loops will be converted to a
    \texttt{gpu.launch} op, the next two loops will be mapped to the warps and
    the remaining compute loops are actually sequential and remain as is.

  \subsection{Completing Latency Hiding}
    In Section~\ref{sec:latency-hiding}, we described latency hiding and
    concluded that it is not complete until we decoupled the loads from stores.
    To achieve this without introducing any
    complexity in the code we first completely unroll the copy loops inside the
    thread block k-loop and then delay the stores such that they happen after
    compute is completed. We show the general structure of the IR that we have
    in Listing~\ref{lst:latency-hiding}. The approach we follow is quite similar
    to the one pointed out in ~\cite{voltaCodegen}.
    \begin{lstlisting}[style=mlir, caption=Gobal memory load latency hiding., captionpos=b,
    label=lst:latency-hiding]
    gpu.launch blocks(%blockIdX, %blockIdY, %blockIdX) in (%arg6 = %c64, %arg7 = %c64, %arg8 = %c1) threads(%threadIdX, %threadIdY, %threadIdZ) in (%arg9 = %c256, %arg10 = %c1, %arg11 = %c1) {
      ...
      %c_reg_0 = gpu.subgroup_mma_load_matrix %C[%26, %27] {leadDimension = 8192 : index} : memref<8192x8192xf32> -> !gpu.mma_matrix<16x16xf32, "COp">
      ...
      // Peeled copy loops for iteration 0 of k-loop.
      scf.for %copy = %c0 to %c4 step %c1 {
	...
      }
      scf.for %copy = %c0 to %c4 step %c1 {
	...
      }
      gpu.barrier
      // Main k-loop
      %res:8 = scf.for %k = %c0 to %c8128 step %c64 iter_args(%c_in_0 = %c_reg_0, %c_in_1 = %c_reg_1 ...) -> (!gpu.mma_matrix<16x16xf32, "COp">, !gpu.mma_matrix<16x16xf32, "COp"> ...) {
	gpu.barrier
	// Global memory loads for iteration i + 1 of k-loop
	%a_next_iter_0 = memref.load %a_cast[%74, %81] : memref<8192x1024xvector<8xf16>>
	%b_next_iter_0 = memref.load %b_cast[%94, %101] : memref<8192x1024xvector<8xf16>>
	...
	scf.for %kk = %c0 to %c64 step %c32 iter_args(%arg16 = %c_in_0, %arg17 = %c_in_1) -> (!gpu.mma_matrix<16x16xf32, "COp">, !gpu.mma_matrix<16x16xf32, "COp"> {
	  ...
	}
	gpu.barrier
	// Shared memory stores for iteration i + 1 of k-loop
	memref.store %b_next_iter_0, %b_smem_cast[%51, %68] : memref<64x17xvector<8xf16>, 3>
	memref.store %a_next_iter_0, %a_smem_cast[%150, %167] : memref<128x9xvector<8xf16>, 3>
	...
      }
      gpu.barrier
      // Peeled compute loop for iteration n-1 of k-loop.
      scf.for %arg14 = %c0 to %c64 step %c32 {
	...
      }
      gpu.subgroup_mma_store_matrix %res#0, %C[%26, %27] {leadDimension = 8192 : index} : !gpu.mma_matrix<16x16xf32, "COp">, memref<8192x8192xf32>
      ...
    }
    \end{lstlisting}

    This is the terminal point for us in terms of optimizations, and this was
    our last step while in the SCF dialect.

  \subsection{Putting It All Together}
    As the previous step was the last one in terms of optimization, the 
    generated IR is now to be set up for execution. There already exists 
    well-designed infrastructure in MLIR to achieve 
    this~\cite{zinenko19gpu,herhut20gpu} and we provide a summary outline of it 
    for completeness.  The existing design in MLIR allows one to represent IR to 
    be executed on an accelerator like the GPU within a single MLIR file. The IR 
    would have two parts: a \texttt{host}-side component, which runs
    on the CPU, and a \texttt{device} side component, or the kernel, which runs
    on the GPU. The host side component invokes the device side component, and
    can wait for its execution to complete or can proceed to do other tasks.
    The lowering paths are slightly different for the host and device
    side components:
    \begin{itemize}
    \item \textbf{Host side compilation:} The host side code is converted to the 
        \texttt{std} dialect and then subsequently to the \texttt{llvm} dialect.
        During conversion to the \texttt{llvm} dialect, operations from the GPU 
        dialect like \texttt{gpu.launch}
        are lowered to function calls to the CUDA driver and CUDA runtime API, 
        through a thin wrapper interface provided by MLIR's CUDA runtime 
        library.  MLIR is then translated to LLVM IR, and target code is 
        generated.  The execution of IR is then made possible through 
        \texttt{mlir-cpu-runner} (using the MLIR's LLVM Orc-based JIT). It takes 
        as arguments the shared libraries to link against, where we can supply 
        the library corresponding to the CUDA driver API.
    \item \textbf{Device side compilation:} The device side code is also
        converted to the \texttt{std} dialect and then to a mix of \texttt{llvm} 
        and \texttt{nvvm} dialect. This is in turn translated to LLVM IR and 
        then to \texttt{PTX} by the \texttt{NVPTX}
        backend in LLVM. \texttt{PTX} is then converted to {\it cubin} (CUDA 
        binary format) using NVIDIA's
        compiler. NVIDIA's compiler is called via the CUDA driver API from MLIR.
        The \texttt{gpu-to-cubin} pass in MLIR has access to the driver API and 
         performs the \texttt{PTX} to cubin compilation and assembly for us. We 
         extended the pass to take this pass to take additional options like 
         optimization level and the maximum number of registers per thread, 
         which are required while compiling \texttt{PTX} to cubin.
    \end{itemize}
    The infrastructure to perform these final steps already existed in MLIR.  
    While our evaluation used the MLIR JIT, ahead-of-time compilation could also 
    be performed with a similar setup. The GPU dialect operations that we 
    introduced, \texttt{gpu.subgroup\_mma\_load\_matrix},
    \texttt{gpu.subgroup\_mma\_store\_matrix} and
    \texttt{gpu.subgroup\_mma\_compute}, were open-sourced and upstreamed to the 
    official LLVM/MLIR repository.

\section{Evaluation}
  In this section, we present the performance of our kernels and compare them with
  CuBLAS~11.2. The evaluation is performed on an NVIDIA Ampere-based Geforce RTX 
  3090 installed on a x86-64 system with an AMD Ryzen Threadripper 3970X CPU, 
  running Ubuntu 20.04 LTS. The following parameters were set for all the 
  experiments:
  \begin{itemize}
    \item The SM clocks were set to the boost frequency mentioned in the
      whitepaper for all the experiments, which is 1695 MHz.
    \item We limit ourselves to statically allocated shared memory, which is
      equal to 48~KB.
    \item The maximum number of registers per thread is set to 255.
  \end{itemize}

  We use NVIDIA Nsight Systems~\cite{nsys} for timing and consider only the 
  kernel runtime for
  calculating the attained TFLOPs. This applies to our kernels as well as
  CuBLAS. We consider different combinations of thread block level tiles and
  warp level tiles and report the best performing version. The reported
  performance has been averaged over ten runs.

  Finally, we consider a matmul of the form \textit{C = AB + C} (all three
  matrices are stored in a row-major layout). We use the \texttt{m16n16k16} 
  version of the WMMA intrinsic and limit ourselves to square problem sizes 
  ranging from 1024 to 16384 with a step of 256. We assume that the problem 
  sizes are multiples of thread block tiles, which are again multiples of warp 
  tiles, which are in turn multiples of the WMMA intrinsic.

  \begin{figure}[!h]
    \centering
    \def\svgwidth{\columnwidth}
    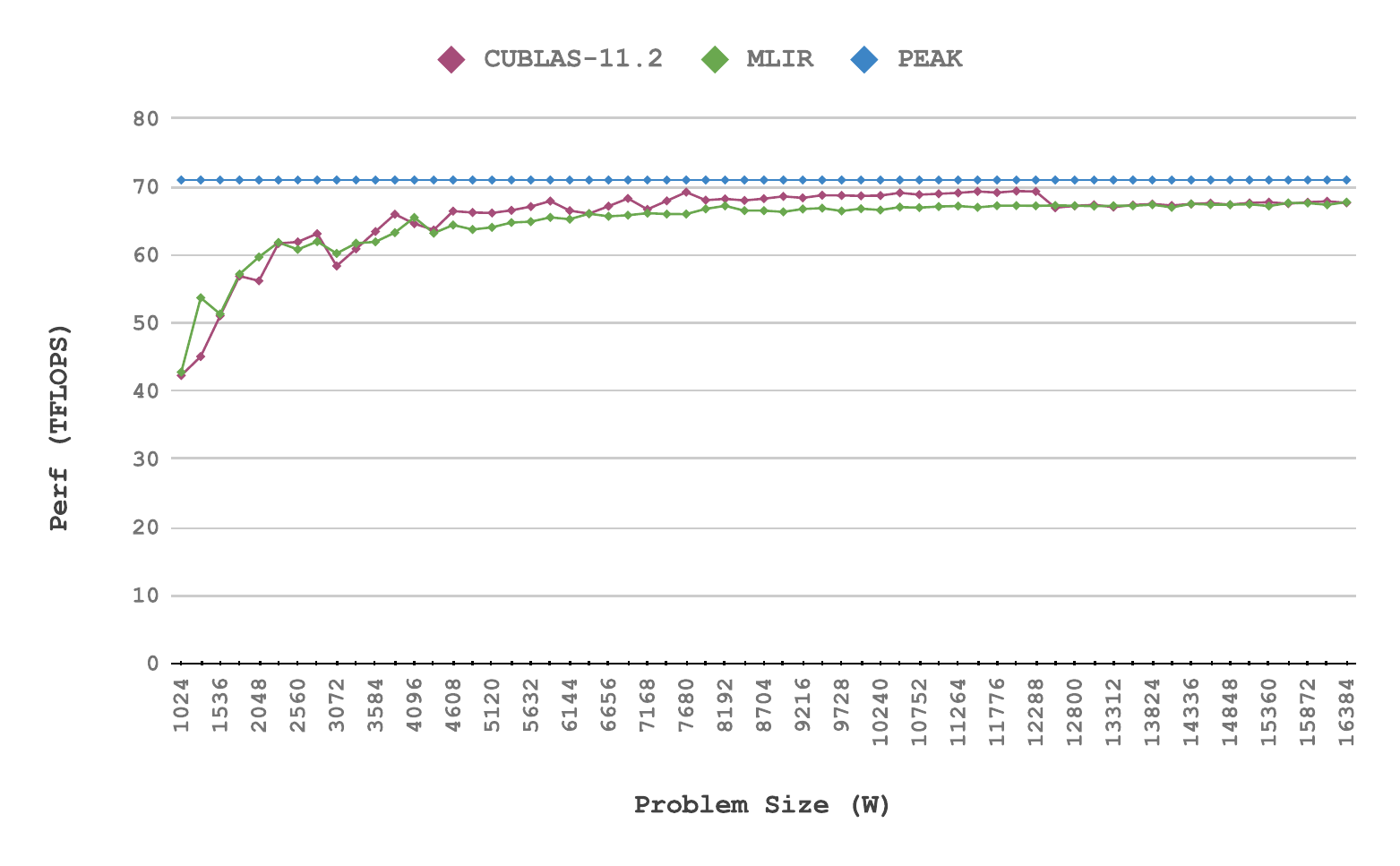
    \caption{Mixed precision (fp16 input, fp32 accumulation and output) 
  performance on square sized
  matrices.~\label{fig:ampere-mixed}}
  \end{figure}

  \subsection{Mixed Precision Performance}
    In this section, we present the performance of our automatically generated 
    mixed precision kernels. Matrix-matrix multiplication having \texttt{A},
    \texttt{B} in \texttt{F16} and the accumulation of the products being done
    in \texttt{F32} is termed as mixed precision matmul. The output matrix
    \texttt{C}, is also in \texttt{F32}. This version is of particular interest
    to us because of its importance in training deep learning 
    models~\cite{mixed-precision-training}.

    We achieve performance that is consistently within 95-119\% of CuBLAS 11.2.
    Comparing our performance with the absolute peak of the device, we sustain 
    95.4\% of the device peak.  Figure~\ref{fig:ampere-mixed} shows the
    performance of our automatically generated kernels on Ampere RTX 3090.
    The figure shows we are very close to CuBLAS on the evaluated sizes. On a 
    few smaller sizes, we outperform CuBLAS. In general, CuBLAS kernels may not 
    be as well-tuned for smaller sizes as they are for the larger ones.  On 
    larger sizes MLIR-generated code is within 2-8\% of cuBLAS performance.  It 
    was further observed that smaller thread block tile sizes like
    \texttt{64$\times$64$\times$64} performed better on smaller problem sizes.
    This indicates that smaller problem sizes are benefited by increased
    occupancy.  While smaller tile sizes reduce reuse for \texttt{A} and
    \texttt{B} in shared memory, they increase occupancy and this benefits small
    problem sizes that may launch relatively fewer thread blocks. On large 
    problem sizes, small tile sizes result in a relatively large number of small 
    threads blocks, and the effect of reduced data reuse becomes more prominent, 
    and increased occupancy does not benefit any more.
    
    \begin{figure}[!h]
      \centering
      \def\svgwidth{\columnwidth}
      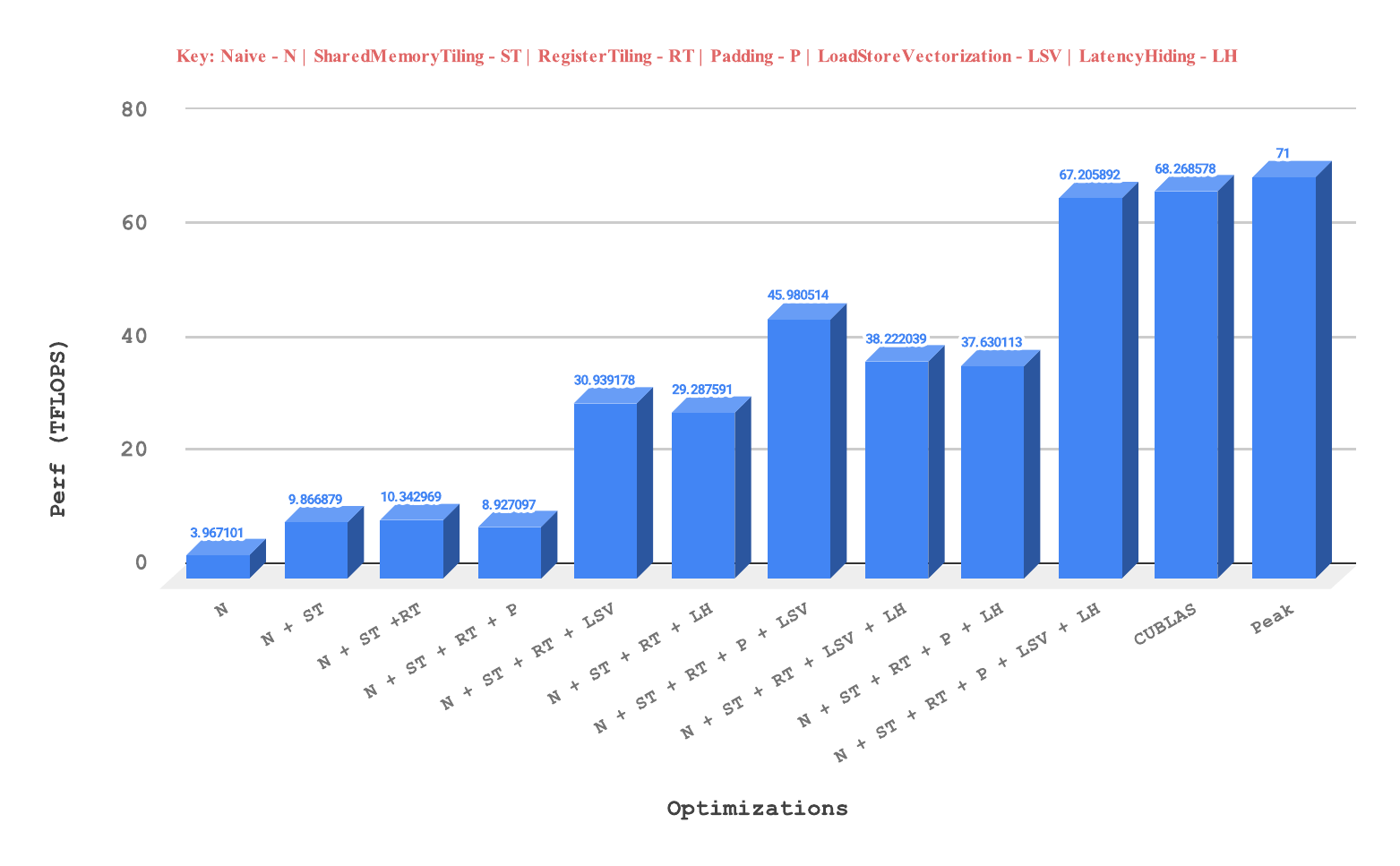
      \caption{Mixed precision performance (fp16 input, fp32 accumulate and 
      output) for M = N = K = 
      8192 with various optimizations enabled and        
      disabled.~\label{fig:gradual-opts}}
    \end{figure}

    An automatic code generation approach also allows us to study the impact of 
    individual optimizations by enabling or disabling optimizations selectively.  
    We show the impact of each of the optimizations discussed earlier in 
    Figure~\ref{fig:gradual-opts} in an incremental manner starting from a naive 
    version to the fully optimized one.

  \subsection{Half Precision Performance}
    \begin{figure}[!h]
      \centering
      \def\svgwidth{\columnwidth}
      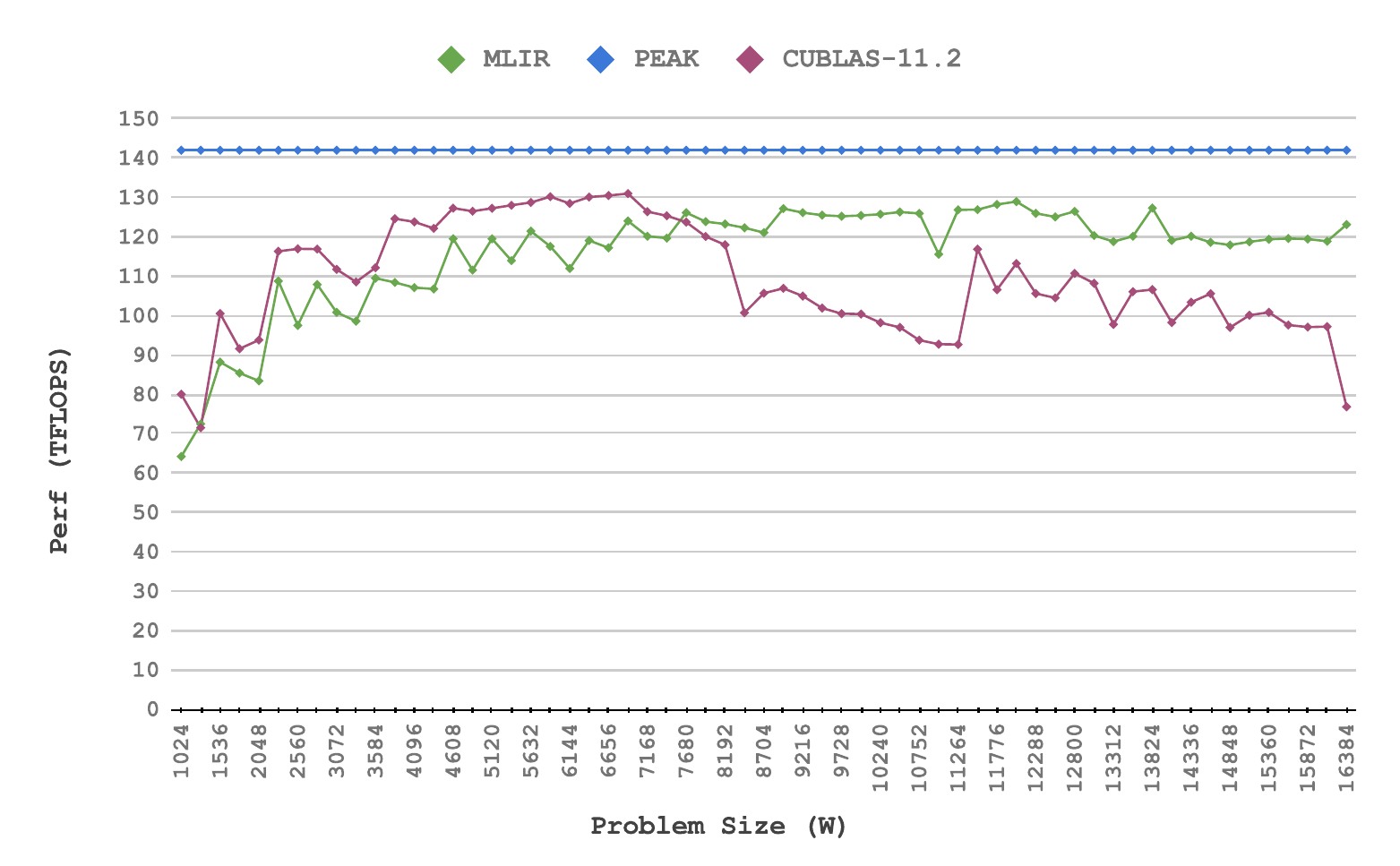
      \caption{Performance with fp16 (input, accumulation and output in fp16) on 
    square sized matrices.
    \label{fig:ampere-half}}
    \end{figure}
    In this section, we present the performance of our automatically generated 
    half precision kernels. In this version of matmul, all three matrices 
    \texttt{A},
    \texttt{B} and \texttt{C} are in \texttt{FP16}. The accumulation of products
    is also done here in \texttt{FP16}. This version is in general faster than
    \texttt{F32} version, but can be prone to imprecision due to a narrower 
    representation for the mantissa and exponent.

    We achieve performance that is consistently 80-160\% of cuBLAS 11.2.
    Figure~\ref{fig:ampere-half} shows the performance of our automatically
    generated kernels on Ampere RTX 3090. We observe that cuBLAS has
    inconsistent performance throughout the range, particularly on problem sizes
    larger than \texttt{W = 8848}.  This suggests that cuBLAS is not well-tuned
    for all problem sizes. In particular, on profiling cuBLAS kernels we observe
    that thread block tile sizes chosen by cuBLAS were actually smaller than the
    ones for which we had optimal performance, for example, for
    \texttt{W=11264}, cuBLAS chooses \texttt{128$\times$128$\times$32}, while we
    choose \texttt{128$\times$256$\times$32}. We have a single stage of
    pipelining for hiding the latency of global memory loads while cuBLAS is
    using five stages.  However, stalls on global memory loads were much more 
    for cuBLAS. This could be as a result of sub-optimal latency hiding.

\section{Conclusion And Future Work}
  We presented early results on automatic code generation targeting specialized
  matmul instructions supported by NVIDIA's tensor cores. These preliminary
  results demonstrate that an automatic code-generator can achieve performance
  competitive to that of hand-tuned libraries in many cases. Experimental 
  results on an NVIDIA Geforce 3090 RTX (based on the NVIDIA Ampere 
  architecture) demonstrated the effectiveness of the presented approach.   
  These results are only meant to serve as a stepping stone for the design of 
  robust code and library generators that are able to not only optimize a single 
  kernel but enable composition and fusion of kernels. This is an area where it 
  is well-known that optimized libraries have limitations. While there have been 
  many efforts to allow fusion and code generation via DSL compilers or graph 
  rewriters, a robust approach based on a unified IR infrastructure has still 
  been missing.

\section{Acknowledgements}
  We thank the MLIR community for the excellent infrastructure that this work
  rests on. We especially thank the developers of Affine, GPU, and LLVM
  dialects for providing useful infrastructure that was used extensively in this 
  work. We also thank other members of the community who reviewed artifacts of 
  this work that were already upstreamed to the official LLVM/MLIR repository.

\end{document}